# High Entropy Alloy CrFeNiCoCu sputtered films: Structure, Electrical Properties and Oxidation

J. Mayandi [1,2], M. Schrade [1,3], P. Vajeeston [4], M. Stange [3], A. M. Lind [3], M. F.Sunding [3], J. Deuermeier [5], E. Fortunato [5], O. M. Løvvik [1,3], A. G.Ulyashin [3], S. Diplas [3,4], P.A. Carvalho [3,6],and T. G. Finstad [1,a)]

[1] Department of Physics, University of Oslo, PO Box 1048, Blindern, N-0316, Oslo Norway
[2] Department of Materials Science, School of Chemistry, Madurai Kamaraj University, India
[3] SINTEF Industry, PO B 124 Blindern, NO-0314 Oslo, Norway.
[4] Department of Chemistry, University of Oslo, PO Box 1048, Blindern, N-0316, Oslo Norway
[5] i3N/CENIMAT, Department of Materials Science, NOVA University Lisbon, Portugal
[6] CeFEMA, Instituto Superior Técnico (IST), Universidade de Lisboa, Lisboa, 1049-001 Portugal

High entropy alloy (HEA) films of CrFeCoNiCu were prepared by sputtering. Their structure was characterized and their electric transport properties studied by temperature dependent Hall and Seebeck measurements. The HEA films show a solid solution with fcc structure. The residual electrical resistivity of the films is around 130 μΩcm which is higher than the Mott limit for a metal while the temperature dependence of the resistivity above 30 K is metal-like but with a small temperature coefficient of resistivity (2 ppm/K). The dominant scattering mechanism of charge carriers is alloy scattering due to chemical disorder in the HEA. The Hall coefficient is positive while the Seebeck coefficient is negative. This is interpreted as arising from an electronic structure where the Fermi level passes through band states having both holes and electrons as indicated by the band structure calculations. Below 30 K the conduction is discussed in terms of weak localization- and Kondo-effects. The HEA structure appears stable for annealing in vacuum, while annealing in an oxygen containing atmosphere causes the surface to oxidize and grow a Cr-rich oxide on the surface. This is then accompanied by demixing of the HEA solid solution and a decrease in the effective resistance of the film.

,

[a)] E-mail: terje.finstad@fys.uio.no

## I. INTRODUCTION

High entropy alloys (HEA) were introduced in 2004 and have attracted a very large attention due to some remarkable material properties reported and the potential to discover materials for applications. [1-5] A typical HEA was made up of five or more elements with near equiatomic concentrations.[6] The entropy of such mixtures was considered as a possibility to stabilize the alloys.[1] The configurational entropy at equiatomic atomic concentrations for quinary alloys is at least 13.5 J/mol K (1.6 R) which can be higher than the entropy of fusion (11–15 J/mol K) for most of the common metals.[7] This led to an interesting feature of HEAs: quite often they were reported to have single-phase or dual-phase solid solution phases (fcc and/or bcc type).[8] Now, the HEA tag may be put on a research field where the phase space around the middle of a multi-element phase diagram is explored even if the material is not fulfilling the original definitions. For some alloys it has been argued that other sources of entropy than the configurational-entropy, such as vibrational -, magnetic - and electronic-entropy, have to be taken into account to explain the stability.[9,10] Also, the presence of secondary phases are common.[11] Here we will stay close to the original meaning of HEA.

HEAs have been synthesized in various ways, but most reports have been on direct casting. [2,12] Yim and Kim [13] gave a survey review of different fabrication processes for HEA such as arc melting, Bridgeman solidification, atomization, laser cladding, powder metallurgy, sputtering, atomic layer deposition and chemical vapor phase techniques. Also melt spinning[2] and laser printing[14] have been reported.

Mechanical properties and dependence on phase stability have been main topics in the literature on HEAs. Mechanical properties of HEAs are complicated[15] and the existing ideas[16] of the mechanisms whereby HEAs are formed and of their properties and production methods are still insufficient for creating a satisfactory clear picture. It has been widely cited that the atoms of HEA are displaced much away from the ideal lattice sites giving local lattice strain and this is a key to many of the observed physical and mechanical properties.[17] There are many simulations succeeding to create that situation with suitable interatomic potentials.[18] However, the hypothesis of large lattice distortions has not been verified satisfactory experimentally, while efforts to measure it by diffuse scattering techniques have recently been reported.[19]

In this work we study the 3d transition-metal HEA CrFeCoNiCu. This HEA has been reported on before.[1,10,16,18,20-29] It may be instructive to remind about the elements in the HEA CrFeCoNiCu and make a qualitative statement about what to expect. The 3d transition metal elements Cr, Fe, Co, Ni and Cu have different crystal structures. They are bcc, bcc, hcp, fcc and fcc respectively. While for the HEA CrFeCoNiCu the crystal structure can be fcc. This fcc alloy then may have a higher stability than the respective elemental systems, while other structures for this system have also been reported on.

For other material systems it is often experienced that results can depend upon preparation method. Thus, it is important to compare results from different laboratories. We report on films made by sputter deposition. For sputter deposited films, in particular, it may be justified to compare results from different laboratories, as even details of the experimental equipment and its condition can make differences. Shaginyan et al.[30] have reported on sputter deposited CrFeCoNiCu alloys, and observed that their films had a nanocrystalline microstructure and crystallize as a two-phase fcc and bcc solid solutions.

Other groups have reported the formation of a single fcc solid solution after deposition.[31-35]

In this work we report on the structure of CrFeCoNiCu films. Knowing that sputter deposition is a non-equilibrium process, which can produce metastable structures, we also test the thermal stability of the structure of the films. We are interested in electron transport in HEAs and will present the temperature dependence of resistivity down to 10 K. This may have direct practical applications as well as a fundamental significance. In a longer perspective we expect a fruitful symbiosis effect between transport theory in metal systems, the possibility of systematic variations in HEA and the increased interest in distortions in HEA studies. Here we reflect over the observation of a high resistivity and apparent metal like character as well as negative low temperature coefficient of resistivity. We report on measurements of the Hall coefficient and Seebeck coefficient which have opposite signs. We explain that by having both electrons and holes at the Fermi surface, and present band calculations that confirms so. The thermal stability of the HEA films under oxidizing ambient is studied next and a surface layer of Cr oxide is observed and accompanied by demixing.

## II. EXPERIMENTAL

The films were deposited on quartz or silica wafer substrates by magnetron sputtering (CVC 601) from a custom-made 8" sputtering target (HR Anlagenbau, GmbH) consisting of equal amounts of the elements Fe, Cr, Co, Ni and Cu. The diffusion pumped base vacuum pressure was about $6\times10^{-5}$ Pa. During the sputtering, the working gas ambience was argon. The flow rate of argon was controlled by a Bronkhurst gas flow controller and set to 30 ml/min. The pressure during deposition was measured with a Baratron capacitance manometer (MKS Instruments) and set to 0.4 Pa through the gate valve with a Swagelok needle valve. A DC power of 200 W was applied. The distance between the target and the substrate was about 6.5 cm, no intentional substrate heating was used during the sputtering, and the substrates were rotated during deposition to obtain lateral uniformity. Several depositions were made with different thickness and deposition rates of 1-3 nm/min. The reported resistivity results are for a thickness of 250 nm.

The films on substrates were then cleaved into small sample pieces (1 cm × 1 cm), and characterized by structure evaluations and transport measurements. Some sample pieces were also annealed in a furnace in flowing $N_2$ (purity 99.999%) $O_2$ or air. Some films were deposited on polymer films and flakes of the film put into a glass capillary for X-ray diffraction (XRD) characterization. We report for a nominal thickness of 250 and 350 nm. We mainly used normal $\theta$-$2\theta$ scans (locked-couple mode) to determine the preferred crystal orientation of the film and the flakes in the capillary was used for identification of compound phases formed in the films using standard instrumentation (Bruker D8 Advance). We were also doing XRD during heating of the samples or flakes to different temperatures. The surface morphology of the films was analyzed by scanning electron microscopy (not shown). X-ray photoelectron spectroscopy (XPS, Kratos AXIS Ultra DLD) with Al $K_\alpha$ X-ray (1486.6 eV) was used to measure elemental composition. Scanning transmission electron microscopy (STEM) with energy dispersive X-ray spectroscopy (EDS) was also used (Titan G2 60–300) to determine the microstructure of the HEA thin films by performing cross-sectional investigations on specimens made by focused ion beam (FIB). The transport characterization consisted of electrical DC Hall and van der Pauw measurements (Lake Shore, 7704 HMS) to observe the temperature dependence of the resistivity and the Hall coefficients giving the dominating carrier type. We used a standard configuration with magnetic field perpendicular to the film surface and four contacts at the sample edges. The Seebeck coefficient was measured on a custom built system[36] using a thermal gradient in the film plane direction. The sign of the Seebeck coefficient also gives the dominating carrier type.

## III. BAND CALCULATION DETAILS

For our structure model construction, we have used primitive FCC Ni as our starting point and the lattice was expanded by 4 X 4 X 4 supercells of nominal composition $Cr_{13}Fe_{13}Co_{13}Ni_{13}Cu_{12}$ and used for the construction of a "special periodic quasirandom structure".[37] In this simulation, the Ni, Fe, Cr, Co and Cu atoms are swapped randomly to find the minimum energy configuration. Total energies of CrFeCoNiCu have been calculated by the projected augmented plane-wave (PAW) implementation of the Vienna *ab initio* simulation package (VASP).[38-41] For the exchange-correlation functional part, we used the Perde-Burke-Ernzerhof (PBE) version of the generalized gradient approximation (GGA) with the Hubbard parameter correction (LDA+U), following the rotationally invariant form [42,43] to perform the structure relaxations and electronic structure calculation.[44]. Effective U values of 4.6, 5.0, 5.1, and 3.7 eV were used for the Fe-, Co-, Ni-, and Cr-*d* states, respectively. The ionic coordinates are fully relaxed until the total energy is changed $10^{-6}$ eV per atom. Ground-state geometries were calculated by minimizing stresses and Hellman-Feynman forces using the conjugate-gradient algorithm with a force convergence threshold of $10^{-3}$ eV Å$^{-1}$. The electronic structure calculation was carried out only for the final, low energy configuration with a 16 X 16 X 16 Γ-centered Monkhorst-pack *k*-point mesh. For all these calculations, a plane-wave cutoff of 600 eV has been used.

## IV. RESULTS AND DISCUSSION

### A.  *Crystal Structure: TEM and XRD results*

Fig.1 shows a STEM cross sectional image from a CrFeCoNiCu film sample. The substrate plane is in the horizontal direction. The electron diffraction from the film reveals an fcc structure (see supplementary material at [URL will be inserted by AIP Publishing] for diffraction pattern). The film consists of columns of the material with the (111) planes parallel to the substrate surface. The lines parallel to the interface in Fig. 1 are twins with coherent (111) Σ3 coincidence site lattice. That result is also consistent with the analysis of the XRD data of the same sample which only gave the (111) diffraction for a 2θ-scan. The lattice parameter was determined to be around $0.3586 \pm 0.0003$ nm which is close the mean, 0.3584 nm, lattice parameter for the elements in that structure. [45]

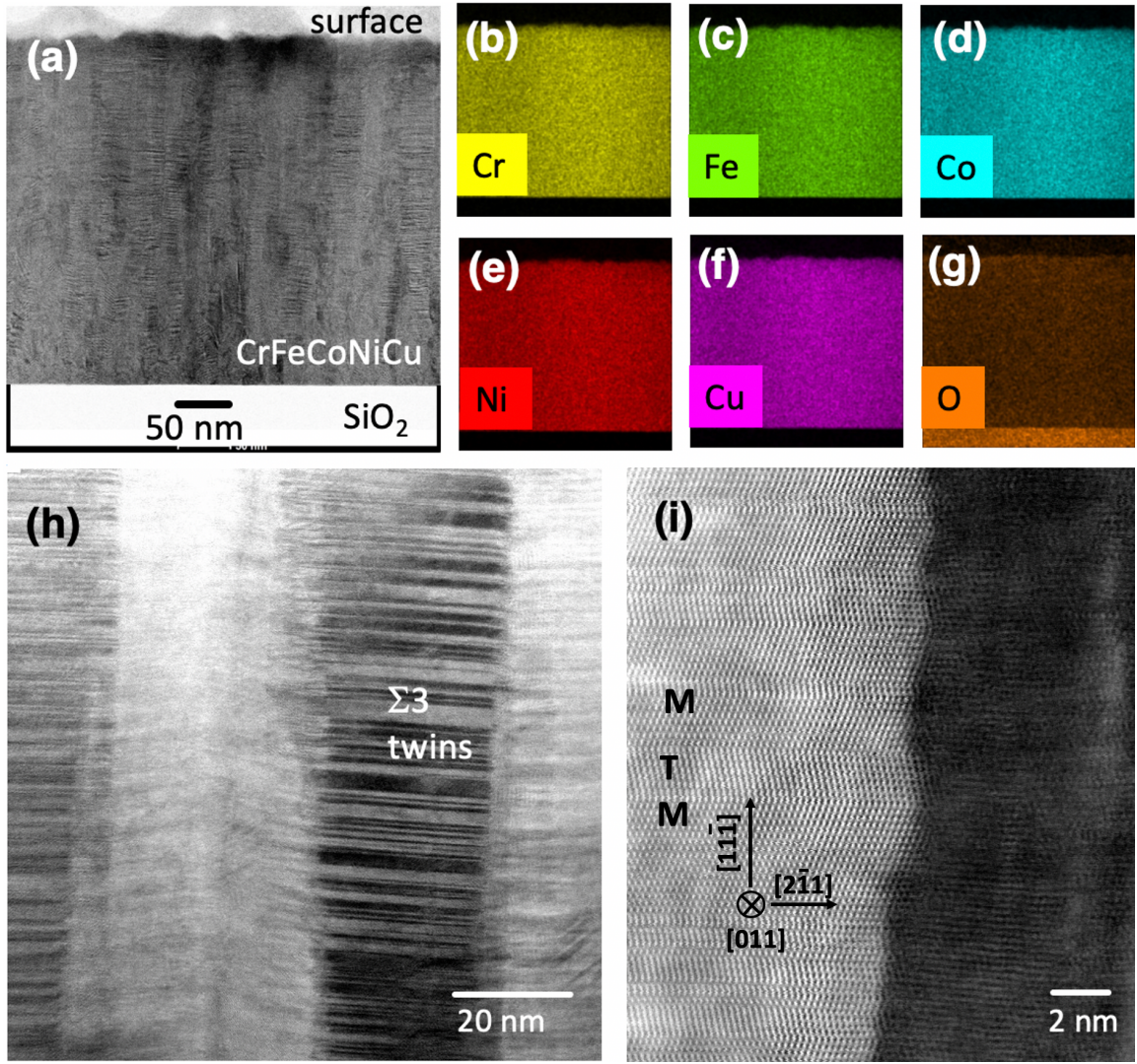

FIG.1. The as-sputtered film of CrFeCoNiCu as imaged by cross sectional STEM. (a) The surface is on top while the SiO$_2$ substrate is at the bottom. Vertical columns are along the growth direction, with horizontal stripes from twin boundaries. (b)-(g) are EDS elemental mapping of the same area as image in (a) for Fe, Cr, Cr, Ni, Cu, and O respectively. In (h) the vertical columns with many twin boundaries is seen. In (i) is shown boundary between two columns. Three twinned regions are indicated by T and M. The crystal directions in region M are by their indices/ The zone axis in M is [011].
.

We also made XRD analysis of the films during heat treatment to check the stability with temperature. In Fig. 2 are patterns obtained from flakes of the film put into a capillary.

The diffraction pattern is consistent with that of an fcc structure indicating that the crystal structure of the sample of Fig.1 is the same as that of Fig.2. The pattern is stable upon heating the sample to 400 °C in vacuum ambient, indicating that the fcc structure is stable up to 400 °C. At that temperature the atomic mobilities are significant, as we will show in section D.1, and new phases should be have been able to form. There are small shifts in the 111 peak ( see Fig. 2(b)). That can be expected from thermal expansion and strain relieve of built in internal strain. The corresponding relative shift in the interplanar distance $d_{111}$ with temperature is $9.8 \times 10^{-6}$ K$^{-1}$. This can be compared to the thermal expansion coefficient of the elements which is 12-13$\times 10^{-6}$ K$^{-1}$ for Co, Ni and Fe and $17 \times 10^{-6}$ K$^{-1}$ for Cu. Thus, the increase in $d_{111}$ with temperature is in the right direction and has a reasonable order of magnitude to be related to thermal expansion also considering relaxation of internal built in strain in the film, which is commonly observed in thin films and is related to the growth process and grain structure.[46-48] We see from Fig 2(b) that the 111 peak after the temperature scan is at a slightly higher value than before. This can be related internal strain release. Most important here is that the fcc structure as such appears stable for annealing up to 400 °C in vacuum. We will see in Section D that the film is not stable for annealing under oxidizing ambient and will be discussed there.

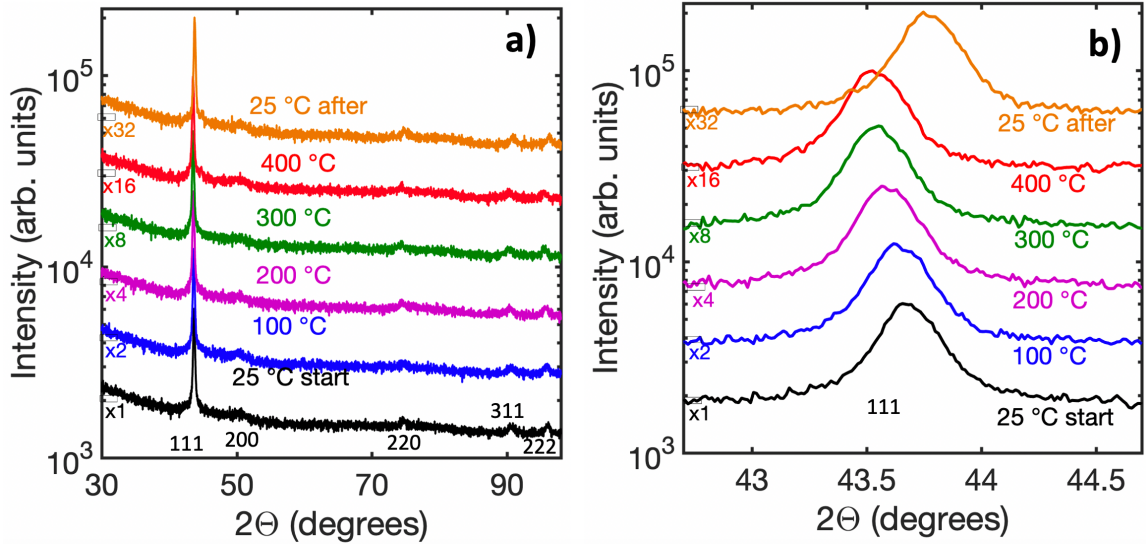

FIG.2. XRD of sputtered film flakes of CrFeCoNiCu measured by Cu-Kα X-rays at increasing temperatures from 25 °C to 400 °C and then after cool down to 25 °C. Each scan were made in 0.5 h before the temperature ramped to the next temperature to stabilize before next scan. (a) All peaks of the scan. (b) The 111 peak.

### 1. Comparison with structure observed by others.

It is interesting to compare the structure reported here with that of the same HEA prepared by different methods. It has often been reported[6,15,20,26,49,50] that material casted from the melt may have inhomogeneities in composition and/or be a mixture of two fcc structures, both with a high concentration of the elements. For example Wang et al.[51] reported two fcc structures for CrFeCoNiCu alloys made by a casting facility attached to an arc melter. The particular structures observed may not necessarily represent an equilibrium state in the phase diagram; it may be a result of the fabrication process, and that applies for the present sputter deposited films as well.

It is also interesting to make comparison with other reports on sputter deposition. A few groups have sputtered the same HEA at room temperature, and reported[31-35] a single fcc solid solution as we observed. Shaginyan et al.[30] reported that their CrFeCoNiCu films had a nanocrystalline microstructure and crystallize as a two-phase fcc and bcc solid solutions. The difference to the present as-deposited case can be related to sputtering conditions, for example our films where prepared with the substrates at room temperature, while those of Shaginyan et al. reached a temperature of 230 °C. We do observe a tendency to formation of some bcc phase under annealing under oxidizing conditions which will be discussed under Section D.1.

## *B. Electrical Characterization*

### *1. Resistivity vs temperature*

The resistivity of the as deposited sample was measured to be 138 μΩcm at room temperature. Figure 3 (a) shows resistivity measured from low temperature to room temperature(RT).

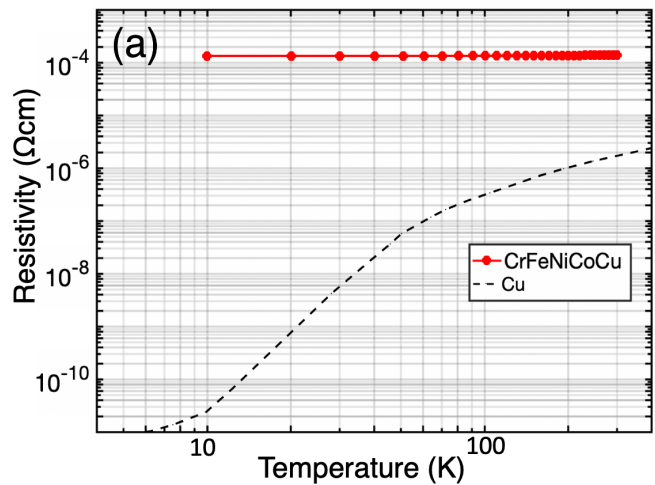

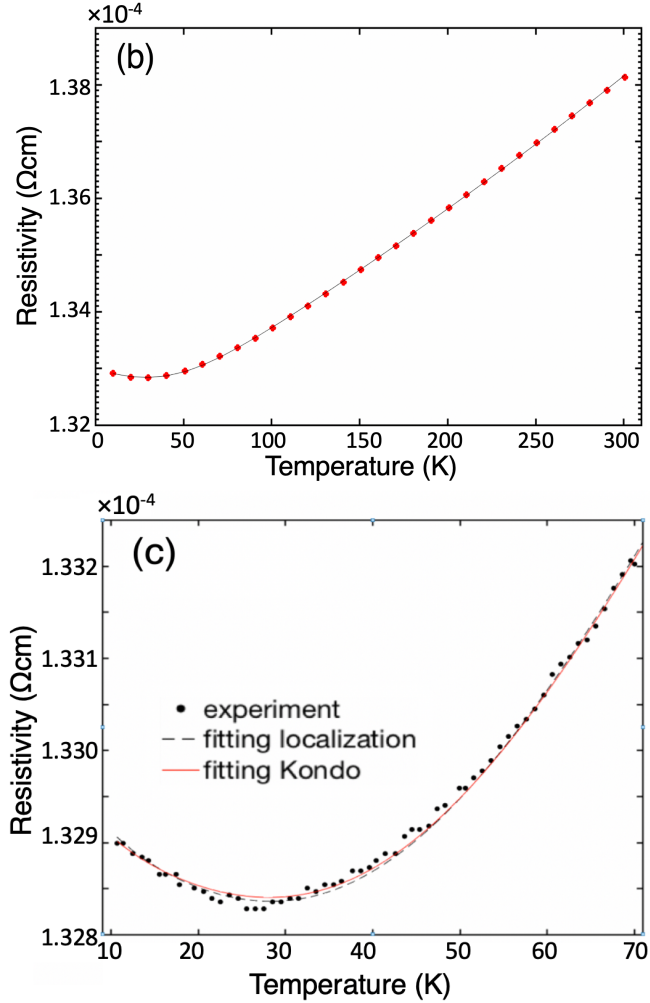

FIG. 3. Resistivity of CrFeCoNiCu measured at different temperatures. a) Red circles are for CrFeCoNiCo, dashed line is for Cu. Note log-log scale  b) Linear scale. Circles are measurements. The line is a fit using Bloch-Grüneisen and electron-electron scattering c) The low temperature part showing the negative temperature coefficient together with two fittings: One adding the Kondo term only and another adding the weak localization term only(see Table I). These fits are indistinguishable.

It is seen that the change in resistivity with temperature is small and that the most important contribution to the resistivity can be described as residual resistivity, $\rho_i$, that is temperature independent and attributed to alloy scattering. Here this scattering comes from the chemical site disorder and local strain created by it. The residual resistance ratio is equal to: RRR=$\rho$(RT)/$\rho$(25 K) = 1.039. The film has a small temperature coefficient of resistance (TCR=∂ln(ρ)/∂T) that is smaller than 2 ppm/K over the whole temperature range. The shape of the temperature variation of the resistivity is seen more clearly in Fig.3(b). It is seen that the resistivity is increasing with temperature. A common basic definition of a metal is a material whose resistivity increases with temperature.[52] It is thus at first natural to treat it as a metal and we will fit the resistivity curve with the following terms appropriate for a metal

$$\rho(T) = \rho_i + \rho_{BG}(T) + \rho_e(T) + \rho_K(T) + \rho_w(T) \quad (1)$$

where $\rho_{BG}(T)$ is due to the phonon electron interactions described by the Bloch-Grüneisen expression, $\rho_e(T)$ is due to the electron-electron interactions or spin interactions induced by the crystal, $\rho_K(T)$ is due to the Kondo effect arising as superposition of spin waves and magnetic atoms and $\rho_w(T)$ is due to weak localization giving quantum interference caused by disorder.

The Bloch-Grüneisen expression is given by [53]

$$\rho_{BG}(T) = 4\mathcal{R}(\Theta_R)\left(\frac{T}{\Theta_R}\right)^5 \int_0^{\Theta_R} \frac{x^5}{(e^x-1)(1-e^{-x})} dx \quad (2)$$

where

$$\mathcal{R}(\Theta_R) = \frac{\hbar}{e^2}\left[\frac{\pi^3(3\pi^2)^{1/3}\hbar^2}{4n_{cell}^{2/3}aMk_B\Theta_R}\right]$$

for a single metal or for an alloy:

$$\mathcal{R}(\Theta_R) = \frac{\hbar}{e^2}\left[\frac{\pi^3(3\pi^2)^{1/3}\hbar^2}{4n_{cell}^{2/3}ak_B\Theta_R}\left(\frac{1}{M}\right)_{ave}\right]$$

where

$$\left(\frac{1}{M}\right)_{ave} = \frac{N_A}{n}\sum_{i=1}^{n}\frac{1}{M_i}$$

Here $\Theta_R$ is the Debye temperature determined by resistivity measurements, $\hbar$ is Planck's constant divided by $2\pi$, $n_{cell}$ is the number of conduction electrons per atom, $M$ is the atomic mass, i.e. atomic weight divided by Avogadro's number $N_A$, $a =$(atomic volume)⅓ , $k_B$ is Boltzmann's constant and $e$ is the elementary charge. For fitting to the experimental data we have used the Padé approximation given by Goetch et al. [53] giving two fitting parameters; $\Theta_R$ and the resistivity at that temperature, $\rho_{\theta_R}$, in which case $\rho_{BG}$ can be written as a function of the normalized temperature $T_n = T/\Theta_R$

$$\rho_{BG}(\rho_{\theta_R}, T_n) = 4.226259 \rho_{\theta_E} T_n^5 \left( \int_0^{\frac{1}{T_n}} \frac{x^5}{(e^x-1)(1-e^{-x})} dx \right) \quad (3)$$

Fig.3(c) shows the low temperature range of measurements together with fits. It is seen that the resistivity has a minimum and then rises with decreasing temperature at the lowest temperatures. The origin of such a resistivity minimum has been the subject of much controversy,[54] and has been associated with quantum interference corrections to electron-electron interactions, weak localization, or to the Kondo effect. A minimum in the $\rho(T)$ curve is a hallmark for the Kondo effect[55], and is very common in all Kondo systems having magnetic impurities showing some kind of disorder. In fact, similar phenomenon of $\rho(T)$ minimum has been observed for many compounds showing spin glass or cluster glass behavior.[56] The Kondo term in equation (1) has the form[57,58]

$$\rho_K(T) = c_m \ln\left(\frac{\mu}{T}\right) \quad (4)$$

where $c_m$ is here a fitting parameter characterizing the strength which is proportional to the exchange integral between interacting spins. The effect of $\mu$, which is a characteristic temperature, is to shift the baseline slightly. For the system at hand this shift is very small compared to $\rho_i$. The term $\rho_w(T)$ in equation (1) has the form[59]

$$\rho_w(T) = r_w \sqrt{T} \tag{5}$$

where $r_w$ here is a fitting parameter. The values of the fit parameters for the best fits of Fig. 3(c) are shown in Table I; There is one fit for setting $r_w=0$, marked "Kondo" and another for setting $c_m=0$, marked "localization" From Fig. 3(c) it is seen that these fits are equally good and one can thus not favor weak localization mechanism over the Kondo effect for giving the negative TCR. It is interesting to compare these measurements with the resistivity of sputtered films of FeCoNiCuGe[60], i.e. Cr is substituted with Ge. The residual resistivity in that case is higher ($\rho_i = 225$ µΩcm) as expected because of the relatively larger difference in potential around the Ge atom. One could then also expect the weak localization effect to get stronger. However, no region with negative TCR was observed at low temperature. That gives some indication that the negative TCR for CrFeCoNiCu can be due to Kondo effect, which is mostly observed with low concentration of magnetic impurities and the Cr tend to couple the other elements antiparallel, impeding large domains, and be similar to systems with a dilute concentration of magnetic impurities.

TABLE I. Three sets of fitting parameters for resistivity of Fig. 3. The zeros are set for each line. $\rho_i$ is the residual resistivity. $\Theta_R$ is the Debye temperature for resistivity, $\rho_{\theta_E}$ is the resistivity at $\Theta_R$. $r_e$ is for electron interactions, $c_m$ is for Kondo effect, $r_w$ is for weak localization.

| $\rho_i$ (Ωcm) | $\rho_{\theta_E}$ (Ωcm) | $\Theta_R$ (K) | $r_e$ (Ωcm /K²) | $c_m$ (Ωcm) | $r_w$(ΩcmK$^{1/2}$) |
|---|---|---|---|---|---|
| 1.3288×10⁻⁴ | 6.03×10⁻⁶ | 391.5 | 1.12×10⁻¹¹ | 0 | 0 |
| 1.3358×10⁻⁴ | 7.0×10⁻⁶ | 363.7 | 2.0×10⁻¹¹ | 0 | 1.37×10⁻⁷ |
| 1.3359×10⁻⁴ | 4.06×10⁻⁶ | 309.9 | 1.83×10⁻¹¹ | 8.9×10⁻⁸ | 0 |

### 2. Regarding the high residual resistivity

Considering that the observed resistivity is very high, we may make some reflections whether it is reasonable to consider it as a metal with metallic conduction (like Cu). We can consider Mott's estimate of the maximum resistivity of a metal: [61]

$$\rho_{max} = \frac{\hbar a}{e^2} \qquad (6)$$

where $\hbar$ is Plank's constant divided by $2\pi$, $e$ the elementary charge and $a$ the distance between atoms. Setting in $a \approx 0.25$nm from the measurement of lattice constant of 0.36 nm by XRD gives $\rho_{max} \approx 127$ µΩcm. There are variations of this criterium, for example the Mott-Ioffe-Regel (MIR) limit[62] essentially replacing $a$ with the inverse of the momentum vector at the Fermi surface $k_F$ giving a similar value for $\rho_{max}$. The observation of resistivities higher than the MIR limit is sometimes taken as an indication that the transport is not governed by quasiparticle transport. Another suggested limit from the uncertainty principle between time and energy for the time between collisions in a quasi-particle description give an estimate given by [62]

$$\rho_{max} = \frac{\hbar}{e^2} \frac{1}{k_F} \frac{k_B T}{E_F} \qquad (7)$$

It appears to us that this relation was introduced to address saturation effects in the increase of resistivity with temperature for materials with a high melting point. In our case, we may have that $E_F \gg k_B T$ in the whole temperature range of measurements, making the limit even lower than the MIR limit.

Nevertheless, the Bloch-Grüneisen expression has been bastardized and applied to situations where its simplified assumptions do not hold, still it provides a fruitful framework for comparison of experimental data. Some of the limits for describing the conduction completely by quasiparticle conduction based upon Bloch-type wave function may not hold, but the validity might be extended and the MIR limit can have a larger deviation from the Mott limit (eq. (6) )if the Fermi surface has considerable pockets around the gamma point.

## *3. Regarding micro structure insensitive resistivity*

Presently we cannot compare directly the resistivity values measured with any published measurements on resistivity of CrFeCoNiCu, neither for bulk material nor thin films. However, we can compare with the bulk resistivity for some similar systems. The room temperature bulk resistivity of Nichrome V is 118 µΩcm, and Kao et al.[63] reported resistivities for bulk homogenous HEA CrFeCoNiAl$_x$ ranging roughly between 100 and 180 µΩcm for different values of *x*. In light of that, the resistivity measured for the present films seems reasonable and within a range to be expected for bulk material. It can be expected that the resistivity of our films is comparable to the bulk value because of the high resistivity. The equivalent mean free path of carriers are of the order of distance between atoms from the Mott limit given by eq. (6).[52] Thus, several of the scattering processes being special for thin films will be unimportant in the present case. The thickness of the present films (250 nm) is obviously much larger than the mean free path, so surface scattering will be insignificant compared to other scattering mechanisms. Also, the mean free path is much smaller than the distance between grain boundaries surrounding the columnar grains and these grain boundaries will not contribute significantly to the resistivity in the present HEA films. Otherwise, grain boundary scattering is an important process for most metals and have been given many theoretical treatments and been studied experimentally by correlating resistivity measurements with detailed microstructure characterization. The motivation behind many studies were the importance of electromigration in metal lines for integrated circuits and also due to it being a mechanism giving critical current limitation in superconductors. In the literature one has reported the resistivity per unit area of grain-boundaries. For fcc metals Al, Cu and Ni which has been reviewed, values of the order of $10^{-16}$ $\Omega m^2$ have been reported.[64-66] That value is actually much smaller than the resistivity per atomic monolayer of the HEA. The present films have many twin boundaries (see Fig. 1) and probably many point defects in the as-deposited state. The twin boundaries are coherent for fcc crystals and their contribution to resistivity is small and are often ignored.[66] The contribution to the resistivity in the HEA is also less than the contribution from a mono layer. Since the observed resistivity is in an expected range for bulk, it indicates that these defects do not dominate the resistivity. Often it is assumed that the defects do not have any temperature dependence in metals.[67] Then these will only contribute to a change in the residual resistivity, $\rho_i$, in Eq.(1). The main contribution to $\rho_i$ comes from alloy scattering, that is due to the chemical disorder of the elements which is also the largest disturbance to a perfect periodic potential of the ideal lattice. It is possible to estimate this by the ab initio method presented by Wu et al.[68] and used for describing trends.

## 4. Seebeck and Hall measurements

Figure 4 shows measurements of the Seebeck coefficient and the Hall coefficient in a range of temperatures. The Seebeck coefficient has a magnitude that is in the lower range of that for metals. We see that the sign of the coefficient is negative. For simple material characterization the sign of the Seebeck coefficient is used as an indication of which carrier types, electrons or holes, are dominating. So, electrons are dominating the Seebeck measurements. For the Hall coefficient we see that its sign is positive, indicating that holes are dominating those measurements. (The sign of the Hall coefficient is actually positive down to 10 K, but we do not have reliable Seebeck measurements at low temperature.) The fact that the Seebeck and Hall coefficients have different signs indicates that we have both electrons and holes at the Fermi level in the material. Both carriers will contribute to these coefficients, but they are weighted with different parameters for the Seebeck coefficient and Hall coefficient. To illustrate that, consider the simplified case that we have only one band/kind of electrons and one kind of holes. Then the Seebeck coefficient $\alpha$ and the Hall coefficient $R_H$ is given by[69]

$$\alpha = \frac{\alpha_p - \frac{\mu_n n |\alpha_n|}{\mu_p p}}{1 - \frac{\mu_n n}{\mu_p p}} \qquad R_H = \frac{r(p\mu_p^2 - n\mu_n^2)}{q(n\mu_n - p\mu_p)^2} \qquad (8,9)$$

where $n$, $\mu_n$ and $\alpha_n$ are the concentration, mobility and Seebeck coefficient respectively and $p$, $\mu_p$ and $\alpha_p$ are those for the holes. The symbol $r$ is the Hall factor which has a value of the order of 1 and can be ignored in the current context. Both $\alpha$ and $R_H$ will change sign when going from only holes to only electrons, but they change sign at different carrier concentrations. Thus, they can have different sign for a certain range of concentrations. We will see in the next section that having both electrons and holes is a reasonable hypothesis for the current case.

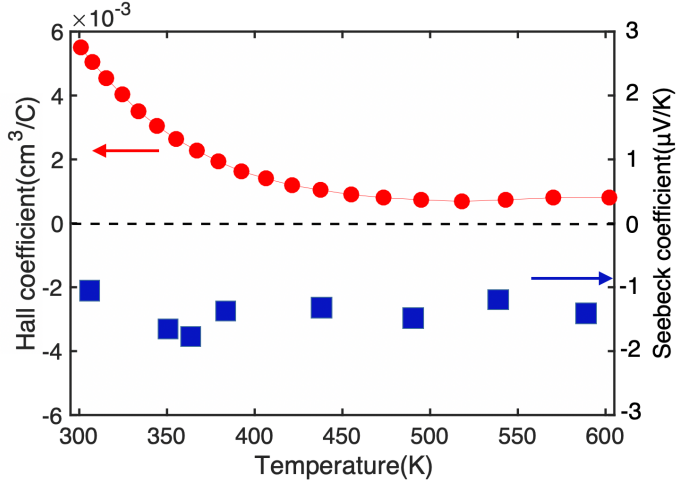

FIG.4. Measurement of CrFeCoNiCu film for various temperatures in the range 300 K to 600 K. The red filled circles show the Hall coefficient, while the blue squares are the Seebeck coefficient. Notice the different sign of these coefficients.

### C.  Band Calculation Results

Figure 5 shows the portion of the band diagram around the Fermi energy $E_F$ calculated as described in section IV. There are many bands that crosses the Fermi level. In particular those bands having pockets around the Fermi level will be states contributing to the conduction. We can identify pockets pointing down as well as pockets pointing up, corresponding to electrons and holes respectively, which is qualitatively in agreement with the hypothesis suggested by the measurements presented in section IV B.4. Making a complete transportation calculation, is considered to be beyond the scope of the work at this stage.

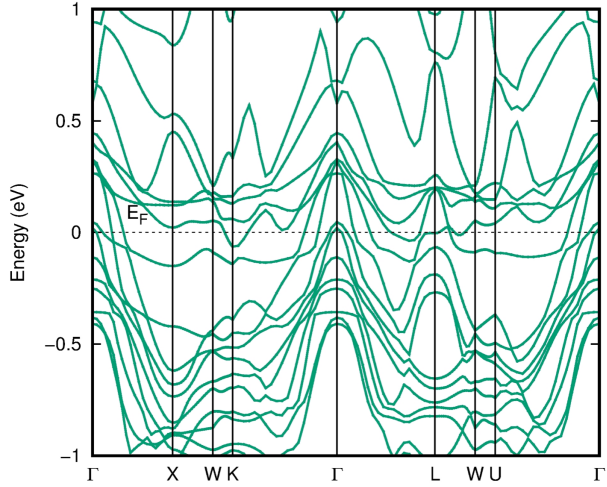

FIG.5. The result of band structure calculation of CrFeCoNiCu . Notice that the Fermi level $E_F$ is intercepted of bands with different curvatures, both positive and negative., indicating that both electrons and holes are at the Fermi level and give contributions the transport

## D.  Thermal stability in oxidizing ambient

Considering that the temperature coefficient of resistivity of the HEA film is very small, it is natural to consider the potential for precision resistor fabrication. For precision resistors it is desirable to have a low TCR. The values we have measured are typically in the range that producers of precision resistors only can deliver on special demand. Sputtering of thin films is also an attractive fabrication procedure for such resistors making automatic trimming by lasers attractive, There are also reports on testing other HEA for resistors[70] giving TCR of -10 ppm/°C. For any practical use of the resistivity of the HEA films it is important to consider their stability with respect to temperature and oxidizing ambient, Therefor we have studied the oxidation of the HEA films. We have oxidized the samples at various temperatures 300, 400, and 500 °C in $O_2$ and in air.

### 1. Structure change by oxidation

Figure 6 shows the changes induced by annealing at 400 °C in $O_2$ for 2h. We see from the STEM cross section that the columnar structure of the as-deposited film remains, but there is a thin, 5nm, layer on the surface. It is seen from the EDS map that the layer is a thin oxide. The same conclusion is arrived at by looking at the XPS signal versus sputtering from the surface in Fig. 7. A thin Cr oxide is seen close to the surface. We find it likely that $Cr_2O_3$ will start to grow on the surface initially and rapidly. This has been observed for Cr thin films[71] and various alloys containing Cr.[72] Out of the binary oxide possibilities, the formation of this oxide gives the largest reduction in free energy. It has been reported for other alloys that the chemical composition of the surface oxide resulting from annealing in oxidizing ambient is strongly dependent on the heating rate.[72] We also observe lattice HRTEM lattice images (see supplementary at [URL will be inserted by AIP Publishing] for images) that indicates a cubic structure, perhaps due to formation of a spinel structure $ACr_2O_4$ (A=Fe, Ni, Co). It is interesting to note that the XPS signals of Fig.7 show mostly Cr and O in the middle of the oxide while a rise of the other metals at the surface. That suggests that the metals diffuse through the grain boundaries of $Cr_2O_3$ and get oxidized at the surface. It has been reported that the growth mode of $Cr_2O_3$ on Cr thin films is by Cr grain boundary diffusion followed by oxidation at the surface.[71]

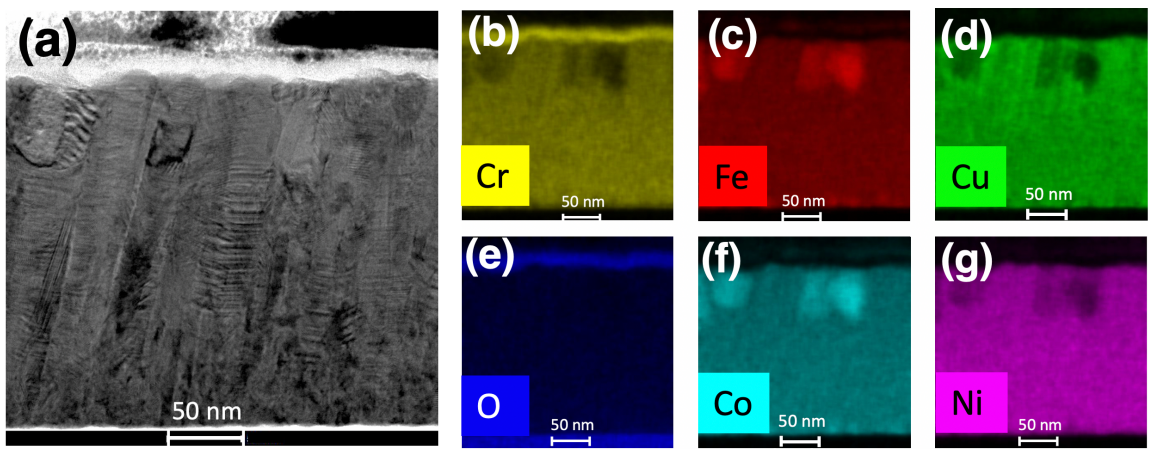

FIG. 6. (a) STEM image of the CrFeCoNiCu film annealed 400 C in $O_2$. The substrate is at the bottom. The bright stripe near the top is an oxide on the surface. The white horizontal stripe at the bottom is the beginning of the substrate.(b)-(g) EDS element maps.

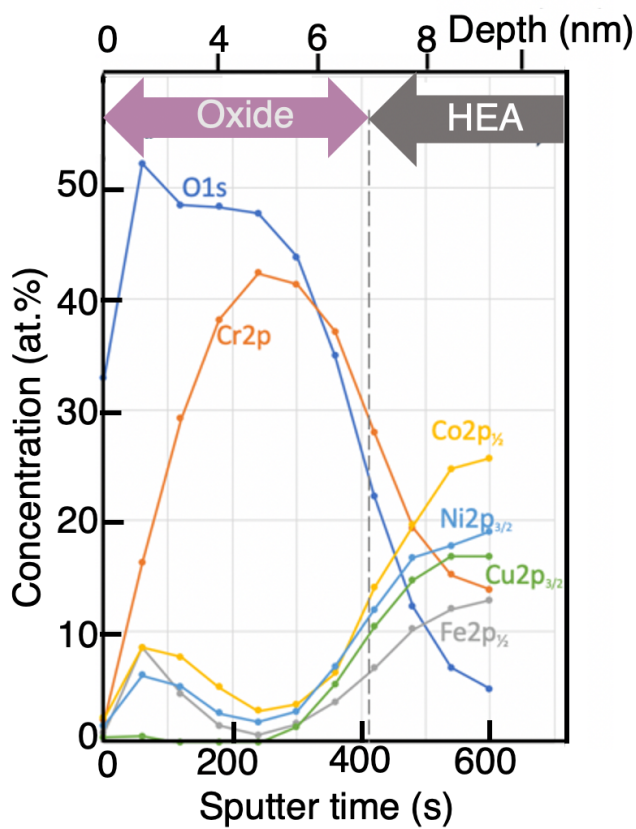

FIG. 7 Element concentration vs depth from surface of CrFeCoNiCu annealed at 400 °C as in Fig 6. Profile created by measuring the XPS signals for electronic transitions

indicated for the elements and normalizing with sensitivity factors to get concentration. The measurements are done after intervals of sputter etching the surface for a time. The sputter time have been converted to depth for convenience by the oxide thickness measured in Fig. 6.

From the EDS element mappings of Fig. 6(b) we also see that the totally mixed HEA film has started to decompose. We see a correlation between the regions enriched in Co and Fe. Those regions are poorer in the other elements. We see that these regions are closer towards the oxide. It is our working hypothesis that the lower content of Cr resulting from the supply of Cr to the surface oxide promotes the demixing observed. We observe more demixing when the film is oxidized at 500 °C for 15 hours, while the oxide layer on the surface has about doubled its thickness, meaning that the oxide grew slowly most of the time compared to the initial growth rate.

## 2. Electrical properties of partly oxidized films

Fig 8 shows the resistivity vs temperature for films that have been oxidized at differnt temperatures (300, 400, 500 °C) in $O_2$ ambient or air as indicated. The as-deposited sample is shown for comparison. We see the general trend that the resistivity decreases with increasing oxidation temperature and the resistivity is lower for those oxidized in $O_2$ than for those oxidized in air. This all scales with the amount of oxidation. We hypothise that it scales with how much Cr has gone out of the HEA film and ended up in the oxide. The oxidation is accompanied by demixing and we will thus measure the resistivity of a film that is inhomogenous. So we will have a composite material, where the resistivity of each part below the oxide will be smaller than the original HEA film due to less chemical disorder. This will yield a lower resistivity for samples exposed to a larger thermal budget during the oxidation. (Annealing out of defects would also contribute in the same direction but would not dominate the reduction as discussed in Section IV B 3). The measured effective resistivity will probably be influenced very little by conduction in the oxide layer, when using van der Pauw measurements for the expected resistivity range of the the oxide.[73]

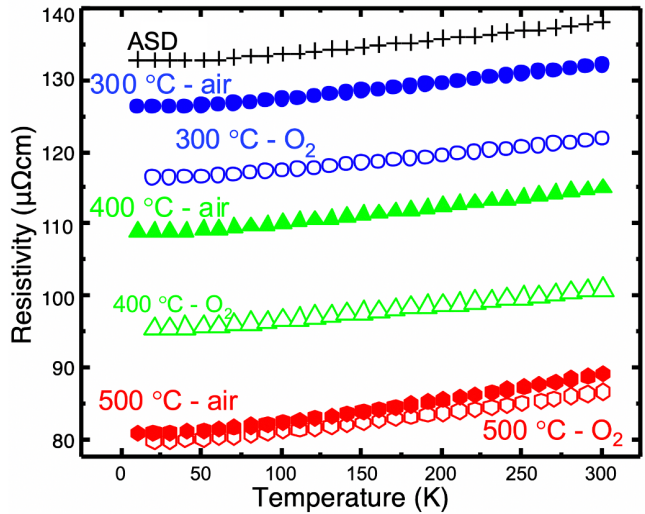

FIG. 8. Resistivity vs measurement temperature for samples oxidized in air or $O_2$ as indicated at the temperature annotated. The curve marked "ASD" is as-deposited.

We also observe that the negative TCR disappears with oxidation. It could be related to less chemical disorder in the films due to demixing. The associated local disorder would also decrease and contribute to the same tendency. Thus, in each region of the film one will have less tendency for weak localization as well. The disappearance of negative TCR could then be due to less weak disorder character, if weak disorder was the primary reason for the negative TCR of the asp sample. However, we believe we cannot rule out that the Kondo effect for the as prepared films is connected to the negative TCR, and the magnetic properties of the films are affected by the oxidation, which is likely considering the ordering occurring by the demixing.

Regarding the stability of the films in terms of applications, it can be rationalized that applications of the resistive properties for high precision are limited to a lower temperature range, probably less than 200 °C. An anneal in an oxidizing ambient will create a surface oxide which will protect the film material. There is a need for high precision resistors for measurements at low temperatures. These often involve magnetic fields or instrumentation that is sensitive to magnetic fields. It is relevant to mention that the as-prepared CrFeCoNiCo films have a magneto-resistance effect that is too small to be measured, meaning the magnetoresistance coefficient is less than $2 \times 10^{-6}$/T. The small magnetoresistance effect may be qualitatively understood by considering the major contribution to the resistivity being disorder, and this disorder is little affected by the magnetic field. For some applications, flexibility, mechanical strength and stability is also important and would have to be addressed considering the particular application.

## V. SUMMARY AND CONCLUSIONS

In this work, we studied sputtered HEA thin films of CrFeCoNiCu by structural and electrical characterization techniques. The crystal structure of the as sputtered thin films is interestingly fcc and the films were homogenous. The resistivity of the films was high, $1.4 \times 10^{-4}$ Ωcm at room which is just above the MIR limit for validity of quasi particle description of the electrical conduction. The dominant contribution to the resistivity comes from chemical and atom position disorder. The temperature dependence of the resistivity still followed a metal description over most of the temperature range 50-300 K. Below 30K the film showed a negative temperature coefficient of resistivity. The temperature coefficient of resistivity was small, which motivated a discussion of practical use for precision resistors and the stability and oxidation properties of the films. A thin layer of Cr rich oxide is grown on the surface of the film under oxidation and this is accompanied by separation of the other elements in the HEA film. Such films may find application at low temperature (<300 °C). For high temperature operation, they may require sealing, or some action to prevent oxidation.

## ACKNOWLEDGMENTS

This work was supported by NFR (Contract No. 275752 HEATER, No.245963 Norfab, No.197405 NORTEM, and No. NN2875k Norwegian supercomputer).

## DATA AVALABILITY

The data that support the findings of this study are available from the corresponding author upon reasonable request.